\documentclass[twocolumn,secnumarabic,showpacs,amssymb, amsmath,tightenlines,aps,prb]{revtex4}
\usepackage{longtable}
\usepackage{hhline}
\usepackage{bm}
\usepackage{epsfig}
\usepackage{graphicx}

\begin{document}

\title[Complex magneto-elastic properties in the frustrated kagome-staircase compounds (Co$_{1-x}$Ni$_x$)$_3$V$_2$O$_8$]
{Complex magneto-elastic properties in the frustrated kagome-staircase compounds (Co$_{1-x}$Ni$_x$)$_3$V$_2$O$_8$}
\author{Q. Zhang$^{1}$, W. Knafo$^{1,2,3}$, P. Adelmann$^{1}$, P. Schweiss$^{1}$, K. Grube$^{1}$, N. Qureshi$^{4,5,6}$, Th. Wolf$^{1}$, and H. v. L\"{o}hneysen$^{1,2}$, C. Meingast$^{1}$}

\address{$^{1}$ Institut f\"{u}r Festk\"{o}rperphysik, Karlsruher Institut f\"{u}r Technologie, 76021 Karlsruhe, Germany.\\
$^{2}$ Physikalisches Institut, Karlsruher Institut f\"{u}r Technologie, 76128 Karlsruhe, Germany.\\
$^{3}$ Laboratoire National des Champs Magn\'{e}tiques Intenses, UPR 3228, CNRS-UJF-UPS-INSA, 143 Avenue de Rangueil,
31400 Toulouse, France.\\
$^{4}$ Institute for Materials Science, University of Technology, D-64728 Darmstadt, Germany.\\
$^{5}$ Institute Max von Laue-Paul Langevin, 38042 Grenoble Cedex 9, France.\\
$^6$ $II$. Physikalisches Institut, Universit\"{a}t zu K\"{o}ln, Z\"{u}lpicherstr. 77, D-50937 K\"{o}ln, Germany.}

\begin{abstract}

High resolution heat capacity and thermal expansion experiments performed on single crystalline kagome-staircase compounds (Co$_{1-x}$Ni$_x$)$_3$V$_2$O$_8$ are presented. The parent compounds Co$_3$V$_2$O$_8$ and Ni$_3$V$_2$O$_8$ undergo a complex sequence of first- and second-order magnetic phase transitions. The low-temperature ($T<40$~K) magnetic entropy evolves monotonously with the doping content $x$, from the full $S=1$ Ni$^{2+}$ magnetic entropy in Ni$_3$V$_2$O$_8$ to half of the $S=3/2$ Co$^{2+}$ magnetic entropy in Co$_3$V$_2$O$_8$. Thermal expansion coefficients $\alpha_i$ ($i = a$, $b$ and $c$) show a strong anisotropy for all (Co$_{1-x}$Ni$_x$)$_3$V$_2$O$_8$ compounds. The low-temperature magnetic distortion indicates that Co-doping (Ni-doping) has similar effects to applying a uniaxial pressures along $a$ or $b$ ($c$). Linear Gr\"{u}neisen parameters $\Gamma_i$ are extracted for the three main axes $i$ and exhibit a complex behavior with both temperature and doping. For each axis, $\Gamma_i$ and $\alpha_i$ exhibit a sign change (at low temperature) at the critical concentration $x_c\simeq0.25$, at which the incommensurate magnetic propagation vector changes. Beyond our study, an understanding of the multiple and complex parameters (magnetic frustration, magnetic anisotropy, mixture of $S=1$ and $S=3/2$ ions, etc.) is now necessarily to bring light to the rich magneto-elastic properties of (Co$_{1-x}$Ni$_x$)$_3$V$_2$O$_8$.

\end{abstract}


\maketitle

\section{Introduction}

The study of the kagome-staircase compounds Co$_3$V$_2$O$_8$ and Ni$_3$V$_2$O$_8$, which have the same orthorhombic (Cmce) crystalline structure, \cite{sauerbrei73} has recently stimulated both experimental and theoretical works. In these systems, two-dimensional and geometrically-frustrated magnetism results from a kagome-staircase arrangement (perpendicularly to the $b$-axis) of the magnetic ions. The magnetic lattice is composed of $S=3/2$ Co$^{2+}$ ions in Co$_3$V$_2$O$_8$ and $S=1$ Ni$^{2+}$ ions in Ni$_3$V$_2$O$_8$. It is made of two non-equivalent magnetic sites. \cite{rogado02} In both compounds, cascades of magnetic transitions have been reported at low temperatures, \cite{chen06,lawes04} which indicates a high number of competing magnetic interactions. Both systems are characterized by a "high temperature" incommensurate (HTI) magnetic order (below $T_{HTI,1}=11.3$~K in Co$_3$V$_2$O$_8$ [\onlinecite{chen06}] and below $T_{HTI,2}=9.1$~K in Ni$_3$V$_2$O$_8$ [\onlinecite{lawes04}]) and by a low temperature commensurate phase (below $T_{FM}=6.2$~K in Co$_3$V$_2$O$_8$ [\onlinecite{chen06}] and below $T_{C'}=2.2$ K in Ni$_3$V$_2$O$_8$ [\onlinecite{lawes04}]). They also have the same easy magnetic axis $\mathbf{a}$. However, their magnetic ground states are associated with different ordering wavevectors. For example, Co$_3$V$_2$O$_8$ is a weak ferromagnet below $T_{FM}=6.2$~K [\onlinecite{chen06}], while Ni$_3$V$_2$O$_8$ is a canted antiferromagnet below $T_{C'}=2.2$~K [\onlinecite{lawes04}]. Interest for these systems was raised with the discovery that Ni$_3$V$_2$O$_8$ enters into a multiferroic regime for $T_C=3.9$~K~$<T<T_{LTI,2}=6.3$~K, where canted antiferromagnetism is accompanied by a ferroelectric polarization. \cite{lawes05} Contrary to Ni$_3$V$_2$O$_8$, Co$_3$V$_2$O$_8$ is not multiferroic. \cite{yasui07} Recent neutron \cite{qureshi08a,qureshi08b} and specific heat \cite{zhang08} studies permitted to draw the doping-temperature magnetic phase diagram of the mixed compounds (Co$_{1-x}$Ni$_x$)$_3$V$_2$O$_8$ and to determine the evolution of the magnetic structure of the system, from Co$_3$V$_2$O$_8$- to Ni$_3$V$_2$O$_8$-type, as a function of doping. From these measurements, only one transition was reported for the samples of nickel concentrations $0.14\leq x\leq0.73$. A change of the magnetic structure was found to occur around the concentration $x_c=0.25$ at which the transition temperature $T_{HTI}\simeq5.5$~K is the smallest. \cite{qureshi08b}

Here, we present a study by specific heat and thermal expansion of single crystals of the mixed compounds (Co$_{1-x}$Ni$_x$)$_3$V$_2$O$_8$ with nickel concentrations $x =0$, 0.12, 0.2, 0.28, 0.35, 0.5, and 1. The same sequences of transitions as those reported in the literature \cite{rogado02,chen06,yasui07,lawes04,wilson07,chaudhury07,kobayashi07} are found for the parent compounds Co$_3$V$_2$O$_8$ and Ni$_3$V$_2$O$_8$. The high resolution of our specific heat and thermal expansion experiments permits us to study the temperature and doping evolution of these transitions in detail and to derive a detailed temperature-doping phase diagram.  Further, magnetic Gr\"{u}neisen parameters, which provide a sensitive probe of the volume/pressure dependence of the magnetic energy scales, can be derived from the combination of thermal expansion and heat capacity data. We will demonstrate that the behavior of the Gr\"{u}neisen parameters in this system is extremely complex, probably due to a high number of competing magnetic energy scales. It exhibits some parallels to the novel behavior expected at quantum phase transitions, \cite{garst05} which will be discussed in detail. This paper is organized as follows. In Section \ref{exp} we provide experimental details. In Section \ref{results} we present the specific heat and thermal expansion results, as well as the calculated uniaxial Gr\"{u}neisen parameters.  In Section \ref{discussion} the phase diagram is derived from our data and the pressure dependences and Gr\"{u}neisen parameters are discussed, and finally the conclusions are presented in Section \ref{conclusion}.

\section{Experimental details} \label{exp}

The (Co$_{1-x}$Ni$_x$)$_3$V$_2$O$_8$ single crystals with Ni-contents $x=0$, 0.12, 0.2, 0.28, 0.35, and 0.5 studied here were grown from fluxes composed of Co$_3$O$_4$, NiO and V$_2$O$_5$. Ni$_3$V$_2$O$_8$ single crystals were grown from a flux composed of BaO, NiO and V$_2$O$_5$. Ni-contents $x$ were determined assuming the validity of the Vegards law, i.e., that the lattice constants (measured using powder x-ray diffraction) vary linearly with $x$. The crystals have been cut to typical dimensions of about $5\times5\times2$~mm$^3$. Specific heat was measured using a Physical Properties Measurement System (PPMS) from Quantum Design; a $^3$He insert was used to reach the lowest temperature of 0.3~K. Since the standard PPMS software is not adapted to the study of sharp transitions, a relaxation method similar to the ones proposed in Ref. [\onlinecite{lashley03,suzuki10}] was used to obtain reliable data near the first- and second-order phase transitions. Linear thermal expansion was measured between 2~K and 300~K using a home-made high-resolution capacitance dilatometer [\onlinecite{meingast90,pott83}] with heating and cooling rates of 6~mK/s or 20~mK/s, within a resolution $\Delta L/L=10^{-8}$. For all samples, thermal expansion was measured along the three main crystallographic directions $\mathbf{a}$, $\mathbf{b}$, and $\mathbf{c}$.

\section{Results} \label{results}

\subsection{Specific heat} \label{heat}

\begin{figure}[b]
    \centering
    \epsfig{file=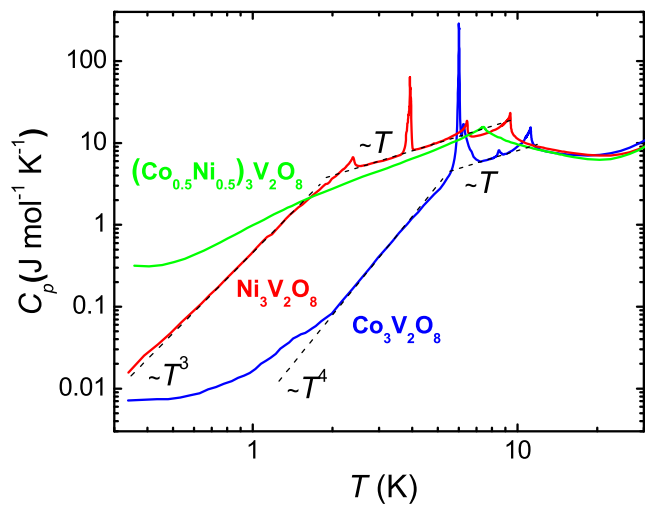,width=85mm}
    \caption{(Color online). Specific heat of Co$_3$V$_2$O$_8$, (Co$_{0.5}$Ni$_{0.5}$)$_3$V$_2$O$_8$ and Ni$_3$V$_2$O$_8$ as a function of temperature in a log-log scale.}
    \label{spec_heat_log_log}
\end{figure}

Specific heat $C_p$ versus temperature $T$ data measured on Co$_3$V$_2$O$_8$, (Co$_{0.5}$Ni$_{0.5}$)$_3$V$_2$O$_8$, and Ni$_3$V$_2$O$_8$ are presented in Fig. \ref{spec_heat_log_log} for $0.3<T<30$~K (log-log scale). The same sequence of transitions as those reported in the literature \cite{chen06,yasui07,lawes04,rogado02} are observed for the two parent compounds Co$_3$V$_2$O$_8$ and Ni$_3$V$_2$O$_8$. Huge first-order anomalies are obtained at $T_{FM}= 6.00\pm0.05$~K in Co$_3$V$_2$O$_8$ and $T_{C}= 3.90\pm0.05$~K in Ni$_3$V$_2$O$_8$, at which $C_p$ reaches 285~Jmol$^{-1}$K$^{-1}$ and 64~Jmol$^{-1}$K$^{-1}$, respectively (almost an order of magnitude higher than in Ref. [\onlinecite{chen06,yasui07,lawes04,rogado02}]). The high quality of our single crystals and an appropriate procedure to analyze the data (relaxation method \cite{lashley03,suzuki10}) permitted to get these high-resolution data. (Co$_{0.5}$Ni$_{0.5}$)$_3$V$_2$O$_8$ has only one second-order phase transition at $7.45\pm0.50$~K (see Appendix \ref{appendixA}). The specific heat follows a $T^3$ power law below 1.5~K in Ni$_3$V$_2$O$_8$  and a $T^4$ power law for $2<T<4$~K in Co$_3$V$_2$O$_8$, leveling off below 2~K. These laws presumably result from low-energy magnetic excitations. The $T^3$ law in Ni$_3$V$_2$O$_8$ is compatible with three-dimensional ungapped antiferromagnetic spin waves, in relation with the canted antiferromagnetic state which sets in below $T_C'\simeq2$~K. Oppositely, the $T^4$ behavior in Co$_3$V$_2$O$_8$ might result from more complex spin dynamics. \cite{notespinwaves} The cascade of magnetic transitions in Co$_3$V$_2$O$_8$ and Ni$_3$V$_2$O$_8$ are also accompanied by an additional signal proportional to $T$. This linear $C_p$ "background" might be related to additional spin fluctuations, similarly to what is observed in heavy-fermion antiferromagnets (see for an example Ref. [\onlinecite{raymond10,lohneysen96}]). For the doped (Co$_{0.5}$Ni$_{0.5}$)$_3$V$_2$O$_8$ compound, the low temperature specific heat is enhanced, indicating stronger low-energy magnetic excitations than in the parent compounds.

\begin{figure}[b]
    \centering
    \epsfig{file=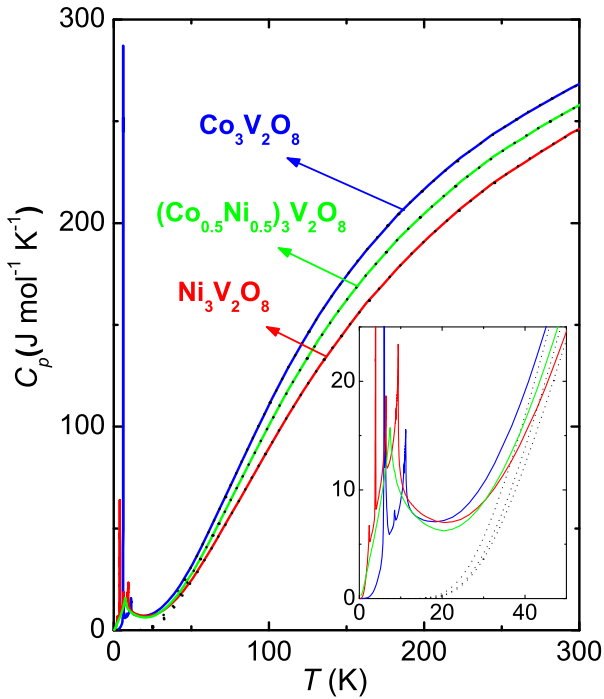,width=80mm}
    \caption{(Color online). Specific heat - and its non-magnetic contribution - versus temperature of Co$_3$V$_2$O$_8$, (Co$_{0.5}$Ni$_{0.5}$)$_3$V$_2$O$_8$ and Ni$_3$V$_2$O$_8$.}
    \label{spec_heat}
\end{figure}

\begin{figure}[t]
    \centering
    \epsfig{file=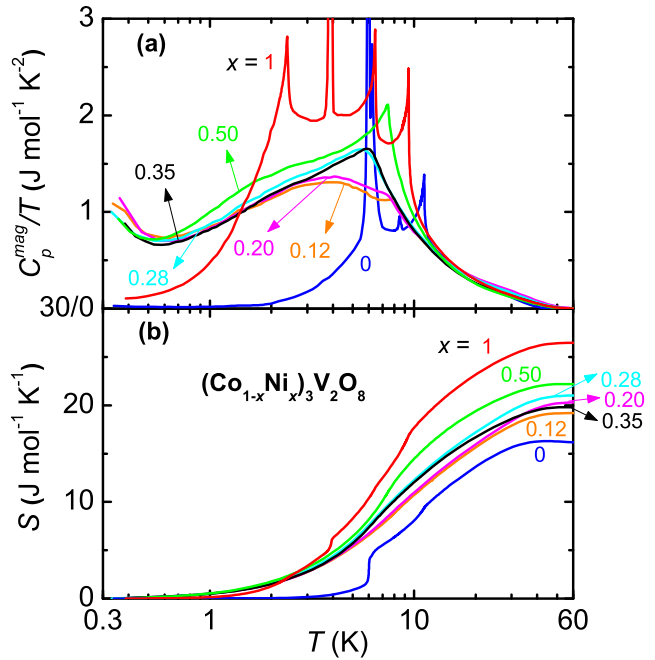,width=85mm}
    \caption{(Color online). (a) Magnetic heat capacity divided by temperature $C_p^{mag}/T$ and (b) magnetic entropy versus temperature of (Co$_{1-x}$Ni$_x$)$_3$V$_2$O$_8$ single crystals.}
    \label{spec_heat_mag}
\end{figure}

Fig. \ref{spec_heat} shows a plot of $C_p$ up to 300~K for Co$_3$V$_2$O$_8$, (Co$_{0.5}$Ni$_{0.5}$)$_3$V$_2$O$_8$, and Ni$_3$V$_2$O$_8$, together with estimates of the phonon background. The non-magnetic signal was modeled by a combination of one Debye and two Einstein phonon modes, which were fit to the data above 60~K (within the approximation that the signal is purely phononic above 60~K). The fitted phonon backgrounds were subtracted from the original data to yield the magnetic heat capacity $C_{mag}$ shown in Fig. \ref{spec_heat_mag} (a) as $C_{mag}/T$ versus $T$, for $0.3<T<60$~K ((Co$_{1-x}$Ni$_x$)$_3$V$_2$O$_8$ samples of concentrations $x=0$, 0.12, 0.2, 0.28, 0.35, 0.5, and 1). From our specific heat data, Co$_3$V$_2$O$_8$ undergoes the series of transitions $T_{HTI,1}=11.26\pm0.02$~K ("high-temperature incommensurate" phase), $T_{HTC,1}=8.50\pm0.05$~K ("high-temperature commensurate" phase), $T_{LTI,2}=6.20\pm0.05$~K ("low-temperature incommensurate" phase), and $T_{FM}=6.00\pm0.05$~K (low-temperature ferromagnetic phase), while Ni$_3$V$_2$O$_8$ is characterized by the transitions $T_{HTI,2}=9.42\pm0.02$~K ("high-temperature incommensurate" phase), $T_{LTI,2}=6.48\pm0.02$~K ("low-temperature incommensurate" phase), $T_{C,2}=3.93\pm0.02$~K (canted antiferromagnetic phase), and $T_{C',2}=2.42\pm0.03$~K (second canted antiferromagnetic phase) - see Appendix \ref{appendixA}.
$T_{HTC,1}$, $T_{LTI,2}$, and $T_{FM}$ in Co$_3$V$_2$O$_8$ and $T_{C,2}$ in Ni$_3$V$_2$O$_8$ are of first-order. \cite{note_order_transition} In contrast to the end compounds, the heat capacities of the mixed crystals with $x=0.12$, 0.2, 0.28, 0.35, and 0.5 show only one phase transition, which is of second order, at $8.19\pm0.10$~K, $7.6\pm0.2$~K, $6.4\pm0.4$, $7.0\pm0.4$~K, and $7.9\pm0.3$~K, respectively. The transition temperature goes through a minimum at $x=0.28$~\%. For the mixed compounds, $C_p^{mag}/T$ increases with decreasing temperature below 0.5~K. Stronger low-temperature magnetic fluctuations might result from a stronger magnetic frustration and could explain the high value of the specific heat at low temperature. Similarly, enhancement of the low-temperature specific heat is generally observed at the quantum phase transition of heavy-fermion systems, \cite{raymond10,lohneysen96} where low-energy fluctuations are known to play a central role. \cite{knafo09}

\begin{figure}[t]
    \centering
    \epsfig{file=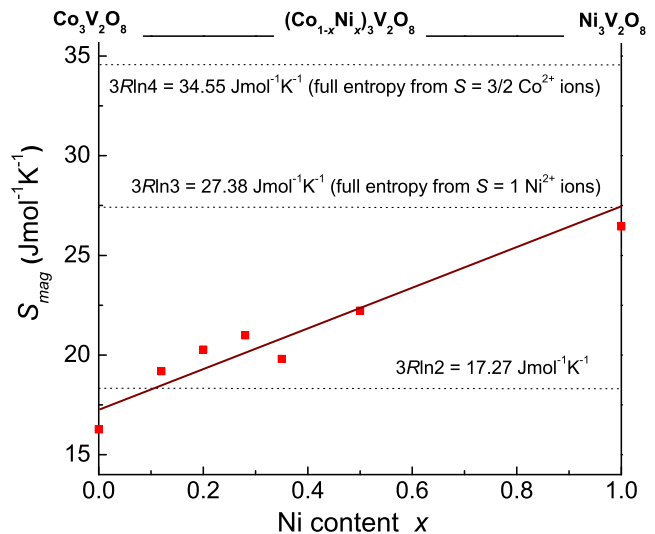,width=85mm}
    \caption{(Color online). Plot of the magnetic entropy of (Co$_{1-x}$Ni$_x$)$_3$V$_2$O$_8$, at $T=40$~K, as a function of the Ni content $x$.}
    \label{entropy}
\end{figure}

The magnetic entropy $S_{mag}$ (from an integration of $C_p^{mag}/T$ versus $T$) is shown for different $x$ in Fig. \ref{spec_heat_mag} (b). Clear step-like increases of the magnetic entropy are observed at the first-order transitions $T_{FM}$ in Co$_3$V$_2$O$_8$ and $T_{C,2}$ in Ni$_3$V$_2$O$_8$. In Fig. \ref{entropy}, the saturated magnetic entropy (at 40 K, from Fig. \ref{spec_heat_mag} (b)) is plotted as a function of $x$. For Ni$_3$V$_2$O$_8$, $S_{mag}$ equals at 40~K the full entropy $3R$ln3 from $S=1$ Ni$^{2+}$ ions, as already reported in Ref. [\onlinecite{rogado02,lawes04}]. For Co$_3$V$_2$O$_8$, $S_{mag}$ reaches at 40~K only half of the full entropy $3R$ln4 expected for the $S=3/2$ Co$^{2+}$ ions, which is also in accordance with previous studies. \cite{rogado02,yasui07} Rogado et al. \cite{rogado02} initially proposed that the missing entropy could be related to additional magnetic transitions occurring below 4~K (4~K was the lowest temperature investigated in their study). We have shown here that the specific heat of Co$_3$V$_2$O$_8$ decreases strongly with decreasing temperature (down to 0.3~K) and the shape of our specific heat data indicates that no further transition might occur below 0.3~K. Alternatively, Yasui et al. \cite{yasui07} proposed that a crystal-field splitting of the $S=3/2$ quadruplet leads to a ground state composed of an effective $S_{eff}=1/2$ doublet, associated thus with a magnetic entropy of $3R$ln2 (half of $3R$ln4). The hypothesis of a stronger crystal field in Co$_3$V$_2$O$_8$ than in Ni$_3$V$_2$O$_8$ is compatible with former susceptibility measurements, \cite{szymczak06,szymczak09,he06} which showed that the magnetic properties of Co$_3$V$_2$O$_8$ are very anisotropic while those of Ni$_3$V$_2$O$_8$ are almost isotropic. We note that a similar situation occurs in the chlorides CoCl$_2$ and NiCl$_2$, where crystal-field effects are stronger in CoCl$_2$ than in NiCl$_2$ [\onlinecite{lines63}].

\subsection{Thermal expansion and Gr\"{u}neisen parameter} \label{th_exp}

\begin{figure}[t]
    \centering
    \epsfig{file=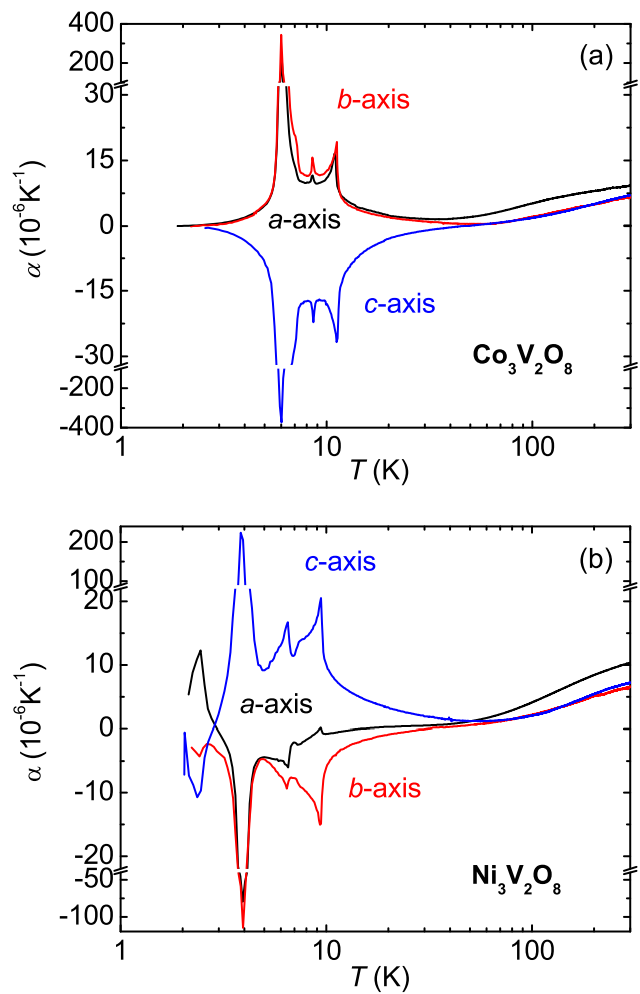,width=85mm}
    \caption{(Color online). Thermal expansion coefficients $\alpha_a$, $\alpha_b$, and $\alpha_c$ versus temperature of (a) Co$_3$V$_2$O$_8$ and (b) Ni$_3$V$_2$O$_8$.}
    \label{alpha_CVO_NVO}
\end{figure}

Fig. \ref{alpha_CVO_NVO} (a) and (b) show the temperature dependence of the thermal expansion coefficients $\alpha_i$ of Co$_3$V$_2$O$_8$ and Ni$_3$V$_2$O$_8$, respectively, for $i=a$, $b$, and $c$ and 2~$<T<$~300~K. They are defined as $\alpha_i=1/L_i\partial L_i/\partial T$, where $L_i$ is the length along $i$. The same sequences of phase transitions as those reported by specific heat (Section \ref{spec_heat}) can be seen in our thermal expansion data for both Co$_3$V$_2$O$_8$ and Ni$_3$V$_2$O$_8$ (see Fig. \ref{alpha_CVO_NVO}, Table \ref{table_transitions}, and Appendix \ref{appendixA}). The thermal expansion anomalies in Co$_3$V$_2$O$_8$ are positive and of roughly the same magnitude for the $a$- and $b$-axes and are negative for the $c$-axis.  Interestingly, this situation is reversed for Ni$_3$V$_2$O$_8$, which has mostly negative effects along the $a$- and $b$-axes and positive anomalies along the $c$-axis.  Above about 50~K, the magnetic contributions vanish and both compounds exhibit very similar expansion coefficients due to the anharmonicity of the phonons.  Here, $\alpha_b$ and $\alpha_c$ are nearly identical for both compounds and $\alpha_a$ is somewhat larger. The similarity of the phonon thermal expansion for both compounds is not unexpected since Co and Ni are neighbors in the periodic table.

\begin{widetext}

\begin{figure}[h]
    \centering
    \epsfig{file=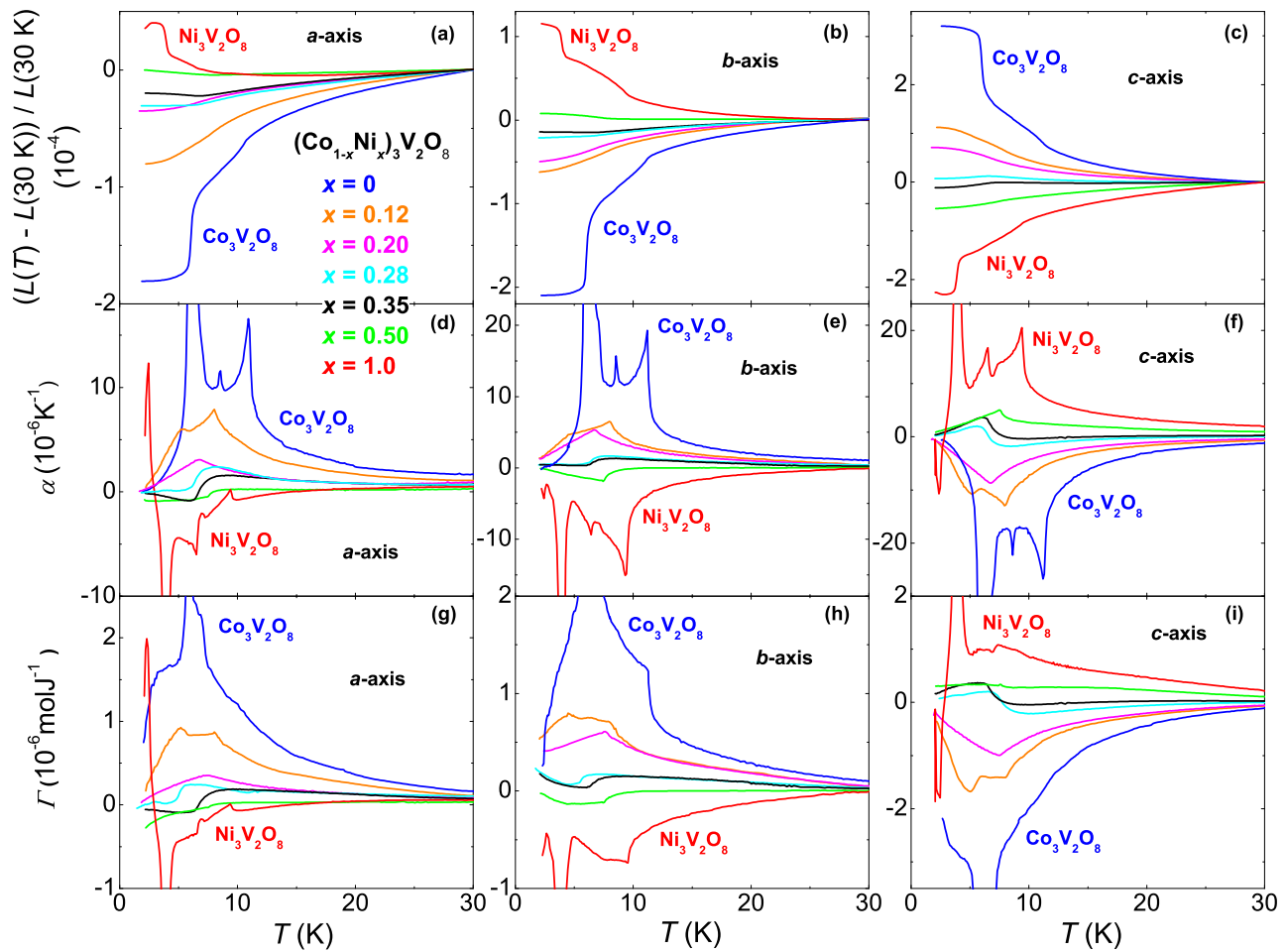,width=170mm}
    \caption{(Color online). Temperature dependence of the relative length $(L(T)-L(\rm{30 K}))/L(\rm{30 K})$ measured along the $a$-, $b$-, and $c$-axes [(a), (b), and (c), respectively], of the thermal expansion coefficient $\alpha$ extracted for the $a$-, $b$-, and $c$-axes [(d), (e), and (f), respectively], and of the Gr\"{u}neisen coefficients $\Gamma$ extracted for the $a$-, $b$-, and $c$-axes [(g), (h), and (i), respectively], for (Co$_{1-x}$Ni$_x$)$_3$V$_2$O$_8$ samples of concentration $x=0$, 0.12, 0.20, 0.28, 0.35, 0.5 and 1.}
    \label{DL_alpha_gamma_all}
\end{figure}

\end{widetext}

The detailed temperature variation of the thermal expansion coefficients, the integrated relative length changes, as well as uniaxial Gr\"{u}neisen parameters are shown in Fig. \ref{DL_alpha_gamma_all} for various Ni concentrations $x$.  We follow Ref. \onlinecite{garst05} in defining the uniaxial Gr\"{u}neisen parameters, $\Gamma_i$, as just the ratio $\alpha_i/C_p$.  The normalized lengths $(L_i-L_i(30~\rm{K})$$/L_i(30~\rm{K})$ for Co$_3$V$_2$O$_8$ and Ni$_3$V$_2$O$_8$ and $i=a,b,c$ again nicely demonstrate the anisotropy discussed earlier and are in good agreement with previously reported values. \cite{chaudhury07,kobayashi07} For the intermediate compounds, there is a progressive change from the Co-type anisotropic behavior to the Ni-type behavior as $x$ increases. The mixed compounds all exhibit only the highest temperature phase transition, except for the $x=0.12$ Ni-sample, which shows two transitions. The anomalies in $\alpha_i$ become much smaller for the intermediate compounds, and the change of behavior from Ni- to Co-like occurs around $x=0.20$-$0.28$.  At these concentrations, the exact behavior is in fact quite strange; above $T_{HTI}$ the data appear to belong to the Co-side, then at $T_{HTI}$ and below $T_{HTI}$ the $\alpha_i$ jumps and sign changes to the Ni-like behavior.  This is most apparent for the $a$- and $c$-axis data (Fig. \ref{DL_alpha_gamma_all} d and f).  We note that this 'jump' occurs only in the thermal expansion data and not in the heat capacity data, where the transition is only marked by a kink in $C_p$, and that, therefore, it is not at all apparent how one can apply the Ehrenfest relation for calculating the pressure dependences of $T_{HT1}$.

In order to gain a better understanding of the above behavior, it is informative to examine the behavior of the magnetic Gr\"{u}neisen parameters shown in Fig. \ref{DL_alpha_gamma_all} (g-i), which look qualitatively similar to the $\alpha_i$ coefficients, but also have some important differences. We first discuss the behavior of $\Gamma_i$ for the end compounds. The first remarkable feature is the complete absence of anomalies in $\Gamma_a$ and $\Gamma_c$ for Co$_3$V$_2$O$_8$ at both $T_{HTI,1}$ and $T_{HTC,1}$; these $\Gamma_i$ coefficients are not constant and both strongly increase in magnitude as the temperature is reduced, peaking at the large first-order phase transition below which the values decrease again. The behavior of $\Gamma_b$ is similar, however a small step is observed in $\Gamma_b$ at $T_{HTI,1}$.  Again, the $\Gamma_i$ coefficients of Ni$_3$V$_2$O$_8$ have opposite signs as those of Co$_3$V$_2$O$_8$, but exhibit qualitatively very similar behavior. Also remarkable is the fact that both $\Gamma_b$ coefficients have a very similar jump-like feature at $T_{HTI}$. The absence of an anomaly in $\Gamma_i$ at these phase transitions implies that the anomalies in $\alpha_i$ and $C_p$ scale using a single parameter, which however is not constant but rather increases with decreasing temperature (Fig. \ref{DL_alpha_gamma_all} (g-i)). The Gr\"{u}neisen parameters probe the volume/pressure dependence of the magnetic energy scales. For the simplest magnetic phase transition, one expects only the exchange energy to be volume/pressure dependent, leading to a temperature independent Gr\"{u}neisen parameter. Clearly our results show that the situation in Co$_3$V$_2$O$_8$ and Ni$_3$V$_2$O$_8$ is much more complicated. Beyond this complexity, the divergent-like behavior of $\Gamma_i$ is reminiscent of the Gr\"{u}neisen parameters of systems close to a quantum phase transition, at which Gr\"{u}neisen parameters necessarily diverge. \cite{paulsen90,garst05}

\begin{figure}[t]
    \centering
    \epsfig{file=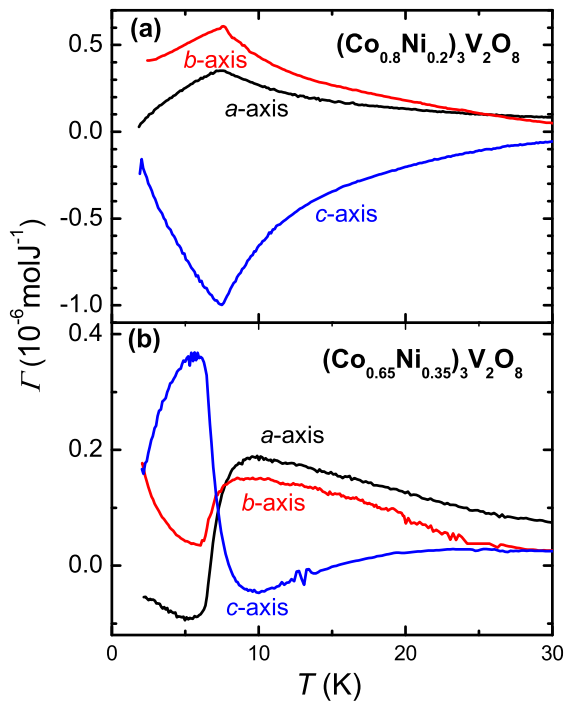,width=75mm}
    \caption{(Color online). Expanded view on the temperature dependence of the Gr\"{u}neisen coefficients of (a) (Co$_{0.8}$Ni$_{0.2}$)$_3$V$_2$O$_8$ and (b) (Co$_{0.65}$Ni$_{0.35}$)$_3$V$_2$O$_8$ for the $a$-, $b$-, and $c$-axes.}
    \label{gamma_doped}
\end{figure}

The Gr\"{u}neisen parameters for the intermediate compounds change, as the $\alpha_i$ coefficients, from the Co- to the Ni-like behavior as the Ni content $x$ increases. Two types of behaviors are particularly interesting and are shown in Fig. \ref{gamma_doped}. For $x=0.2$, all $\Gamma_i$ coefficients appear to diverge in magnitude down to $T_{HTI,1}$, where there is a sharp kink in the curves and below which the magnitude decreases again sharply. This behavior is clearly different from the pure Co$_3$V$_2$O$_8$ case, where the "HTI$_1$" transition is hardly visible in $\Gamma_i$. However, the overall behavior looks like what one would obtain if all the transitions occurring in Co$_3$V$_2$O$_8$ were replaced by a single second-order transition. For $x=0.35$, on the other hand, all $\Gamma_i$ exhibit a jump-like change at $T_{HTI,2}$, with clear sign changes of $\Gamma_i$ occurring for the $a$- and $c$-axes.

All the above features point to the very complex nature of magnetism in (Co$_{1-x}$Ni$_x$)$_3$V$_2$O$_8$, which results from an unusual combination of geometrical frustration, a complex set of competing magnetic interactions, possible disorder, as well as the doping-induced change of the ion moment (from $S=1$ to $S=3/2$) etc.

\begin{figure}[b]
    \centering
    \epsfig{file=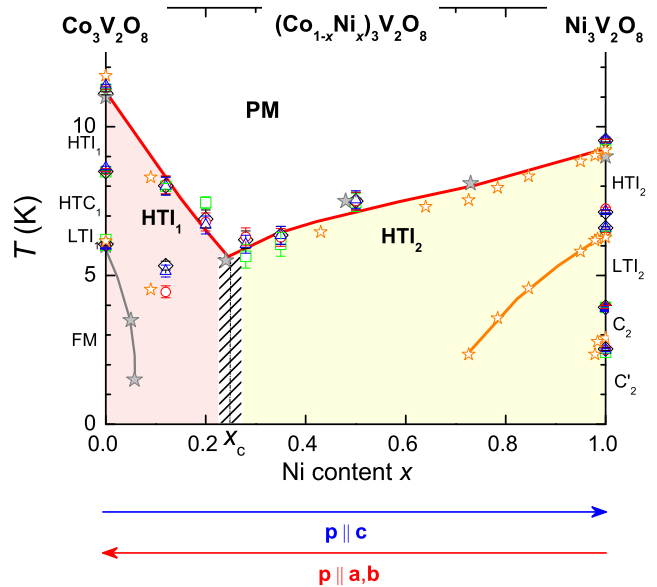,width=85mm}
    \caption{(Color online). Magnetic ($T$,$x$) phase diagram of (Co$_{1-x}$Ni$_x$)$_3$V$_2$O$_8$ extracted from our specific heat (green squares) and thermal expansion measurements (black diamonds: $a$-axis, red circles: $b$-axis, and blue triangles: $c$-axis). Full grey stars corresponds to transition temperatures extracted from neutron experiments by Qureshi et al. \cite{qureshi08a,qureshi08b} and open orange stars to transitions obtained from specific heat by Kumarasiri and Lawes. \cite{kamarasiri11} Arrows indicate that uniaxial pressures applied along $a$ and $b$ are equivalent to Co-doping, while a uniaxial pressure applied along $c$ is equivalent to Ni-doping.}
    \label{phase_diag}
\end{figure}

\section{Phase diagram, pressure dependences, and discussion} \label{discussion}

Fig. \ref{phase_diag} presents the doping - temperature phase diagram of (Co$_{1-x}$Ni$_x$)$_3$V$_2$O$_8$ extracted from specific heat and thermal expansion data (see Section \ref{results} and Appendix \ref{appendixA}) and from neutron scattering experiments. \cite{qureshi08b} The different sets of data agree well and indicate that $T_{HTI}$ is minimal at a critical concentration $x_c\simeq0.25$. The antiferromagnetic phase is noted HTI$_1$ for $x<x_c$ and HTI$_2$ for $x>x_c$ (with the transition temperatures $T_{HTI,1}$ and $T_{HTI,2}$, respectively). At low temperature, a first-order phase transition occurs at the critical concentration $x_c$, where the direction of the incommensurate propagation vector changes. \cite{qureshi08b} Changes of signs in the thermal expansion and Gr\"{u}neisen coefficients are observed close to $x_c$. Additionally, an enhancement of the magnetic specific heat indicates the presence of strong magnetic fluctuations at low temperature, which probably result from competing magnetic interactions.

The transition temperatures extracted here for the compounds (Co$_{1-x}$Ni$_x$)$_3$V$_2$O$_8$, as well as their order and the sign of their uniaxial pressure dependences are summarized in Table \ref{table_transitions} (see also Appendix \ref{appendixA}). Using the Clapeyron relation:
\begin{eqnarray}
\frac{\partial T_x}{\partial
p_i}=\frac{\Delta L_iV}{\Delta S},
    \label{clapeyronrelation}
\end{eqnarray}
where $\partial T_x/\partial p_i$ are the uniaxial pressure dependences of a first-order transition temperature $T_x$, $\Delta L_i$ and $\Delta S$ the length and entropy changes, respectively, $p_i$ a uniaxial pressure applied along $i$ and $V$ the molar volume, we calculate the uniaxial pressure dependences of $T_{HTC,1}$ and $T_{FM}$ in Co$_3$V$_2$O$_8$ and $T_{C}$ in Ni$_3$V$_2$O$_8$ (see Table \ref{table_first_order}). We find that $\partial \rm{ln}$$T_{C}/\partial p_i\simeq-5\partial\rm{ln}$$T_{HTC,1}/\partial p_i\simeq-5/2\partial \rm{ln}$$T_{FM}/\partial p_i$. The signs of these pressure dependences indicate that a uniaxial pressure applied along $a$ or $b$ induces an increase of $T_{HTC,1}$ and $T_{FM}$ and a decrease of $T_{C}$, while a uniaxial pressure applied along $c$ has the opposite effects. The uniaxial pressure dependences of a second-order transition temperature can be extracted using the Ehrenfest relation:
\begin{eqnarray}
\frac{\partial T_x}{\partial
p_i}=\frac{\Delta\alpha_iVT_x}{\Delta C_p},
    \label{ehrenfestrelation}
\end{eqnarray}
where $\Delta\alpha_i$ and $\Delta C_p$ are the thermal expansion and specific heat variations, respectively, induced at the transition. Table \ref{table_second_order} presents the pressure dependences extracted for the second-order transitions $T_{HTI,1}$ in Co$_3$V$_2$O$_8$ and $T_{HTI,2}$, $T_{LTI,2}$, and $T_{C'}$ in Ni$_3$V$_2$O$_8$. The pressure dependences of $T_{HTI,1}$ and $T_{LTI,2}$ have the same signs as that of the first-order transitions from the same compounds. $T_{HTI,2}$, $T_{x,2}$, and $T_{C'}$ can be considered anomalous, since their uniaxial pressure dependences have different signs from that of the first-order transition $T_{C}$.

\begin{widetext}

\begin{table}[h]
\caption{Values of the transitions temperatures, orders of the transitions, and signs of the uniaxial pressure dependences along $a$, $b$, and $c$ of the transition temperatures.}
\begin{ruledtabular}
\begin{tabular}{cllllll}
Ni-content&Transition&$T_x$&Order&Sign of& Sign of& Sign of\\
$x$&temperature&(K)&&$\partial T_x/\partial p_a$&$\partial T_x/\partial p_b$&$\partial T_x/\partial p_c$\\
\hline
&$T_{HTI,1}$&$11.25\pm0.10$&second&$+$&$+$&$-$\\
0&$T_{HTC,1}$&$8.55\pm0.10$&first&$+$&$+$&$-$\\
(Co$_3$V$_2$O$_8$)&$T_{LTI,1}$&$6.2\pm0.1$&first&&&\\
&$T_{FM}$&$6.00\pm0.05$&first&$+$&$+$&$-$\\
\hline
0.12&$T_{HTI,1}$&$8.0\pm0.3$&second&$+$&$+$&$-$\\
&$T_{L}$&$5.0\pm0.5$&second&$+$&$+$&$-$\\
\hline
0.20&$T_{HTI,1}$&$7.0\pm0.5$&second&$+$&$+$&$-$\\
\hline
0.28&$T_{HTI,2}$&$6.0\pm0.4$&second&$-$&$-$&$+$\\
\hline
0.35&$T_{HTI,2}$&$6.3\pm0.4$&second&$-$&$-$&$+$\\
\hline
0.50&$T_{HTI,2}$&$7.5\pm0.3$&second&$-$&$-$&$+$\\
\hline
&$T_{HTI,2}$&$9.5\pm0.1$&second&$+$&$-$&$+$\\
1&$T_{x,2}$&$7.20\pm0.05$&second&$+$&$+$&$-$\\
(Ni$_3$V$_2$O$_8$)&$T_{LTI,2}$&$6.5\pm0.1$&second&$-$&$-$&$+$\\
&$T_{C}$&$3.90\pm0.05$&first&$-$&$-$&$+$\\
&$T_{C'}$&$2.45\pm0.10$&second&$+$&$-$&$-$\\
\end{tabular}
\end{ruledtabular}
\label{table_transitions}
\end{table}

\end{widetext}

The magnetic transitions in the mixed compounds are not associated with well-defined step-like anomalies in $S$, $L_i$, $C_p$ or $\alpha_i$, and an extraction of their uniaxial pressure dependences cannot be done as for the parent compounds. We will limit here the discussion to the signs of the uniaxial pressure dependences of $T_{HTI}$ (summarized in Table \ref{table_transitions}), which are given by the signs of the thermal expansion anomalies at $T_{HTI}$. The trend indicated by the first-order transitions in the parent compounds is followed by the doped compounds. Indeed, $\partial T_{HTI,1}/\partial p_a,\partial T_{HTI,1}/\partial p_a>0$ and $\partial T_{HTI,1}/\partial p_c<0$ on the Co-side($x<x_c$) while, $\partial T_{HTI,2}/\partial p_a,\partial T_{HTI,2}/\partial p_a<0$ and $\partial T_{HTI,2}/\partial p_c>0$ on the Ni-side($x>x_c$). Since $T_{HTI,1}$ decreases with $x$ while $T_{HTI,2}$ increases with $x$ (cf. phase diagram of (Co$_{1-x}$Ni$_x$)$_3$V$_2$O$_8$ in Fig. \ref{phase_diag}), a first approximation is that the effects of a uniaxial pressure along $c$ are similar to that of Ni-doping, while the effects of a uniaxial pressure along $a$ or $b$ are equivalent to that of Co-doping. This simple picture is compatible with the signs of most of the pressure dependences of the parent compounds, i.e., for all the first- and second-order transitions of Co$_3$V$_2$O$_8$, and for the transitions $T_{LTI,2}$ and $T_{C}$ in Ni$_3$V$_2$O$_8$. $T_{HTI,2}$, $T_{x,2}$, and $T_{C'}$ in Ni$_3$V$_2$O$_8$ are exceptions to the picture proposed here, since their pressure dependences are not consistent with it.

To understand the magnetic properties of (Co$_{1-x}$Ni$_x$)$_3$V$_2$O$_8$, a microscopic description of the magnetic interactions and of their uniaxial pressure dependences is now needed (cf. Y$_{1-x}$La$_x$TiO$_3$, where uniaxial pressure dependences of $T_C$ and $T_N$ were related to distortion-induced modifications of the super-exchange [\onlinecite{knafo09b}]). However, the high number of parameters is such that solving the (Co$_{1-x}$Ni$_x$)$_3$V$_2$O$_8$ problem appears to be one of the most challenging puzzles in modern magnetism. Indeed, both parents compounds are subjects to a cascade of magnetic transitions below 12 K, for which commensurate and incommensurate wavevectors have been reported, indicating the presence of a high number of magnetic interactions. Magnetic frustration, but also low-dimensional magnetism, are key ingredients to understand the magnetic properties of this family of compounds. The fact that Ni$_3$V$_2$O$_8$ is made of $S=1$ ions, while Co$_3$V$_2$O$_8$ is made of $S=3/2$ ions, affects surely the magnetic interactions and the magnetic single-ion anisotropy. It should be considered to understand the properties of the parent compounds, as well as that of the mixed compounds. The kagome-staircase lattice of the system is also made of two non-equivalent magnetic sites. Finally, a consequence of this high number of complex magnetic properties, the presence of a multiferroic phase in  Ni$_3$V$_2$O$_8$, has still an enigmatic origin and will certainly deserve again a lot of efforts before being understood.

\begin{widetext}

\begin{table}[h]
\caption{Step-like variations of the entropy $\Delta S$ and the relative lengths $\Delta L_a/L_a$, $\Delta L_b/L_b$, and $\Delta L_c/L_c$ at the first-order transition temperatures $T_{HTC,1}$ and $T_{FM}$ in Co$_3$V$_2$O$_8$ and $T_{C}$ in Ni$_3$V$_2$O$_8$, and uniaxial pressure dependences of these transitions extracted using the Clapeyron relation.}
\begin{ruledtabular}
\begin{tabular}{cllllllll}
Ni-content&Transition&$\Delta S$&$\Delta L_a/L_a$&$\Delta L_b/L_b$&$\Delta L_c/L_c$&$\partial(\rm{ln}$~$T_x)/\partial p_a$&$\partial\rm{ln}$~$T_x/\partial p_b$&$\partial\rm{ln}$~$T_x/\partial p_c$\\
$x$&temperature&(mJ/molK)&&&&(kbar$^{-1}$)&(kbar$^{-1}$)&(kbar$^{-1}$)\\
\hline
0&$T_{HTC,1}$&$0.04\pm0.01$&$0.004\pm0.001$&$0.009\pm0.002$&$-0.010\pm0.002$&$0.11\pm0.05$&$0.21\pm0.10$&$-0.26\pm0.11$\\
(Co$_3$V$_2$O$_8$)&$T_{FM}$&$3.19\pm0.30$&$0.54\pm0.07$&$0.80\pm0.10$&$-1.13\pm0.15$&$0.24\pm0.05$&$0.36\pm0.08$&$-0.51\pm0.12$\\
\hline
1 (Ni$_3$V$_2$O$_8$)&$T_{C}$&$1.08\pm0.10$&$-0.26\pm0.03$&$-0.34\pm0.04$&$0.705\pm0.080$&$-0.51\pm0.11$&$-0.67\pm0.14$&$1.41\pm0.29$\\
\end{tabular}
\end{ruledtabular}
\label{table_first_order}
\end{table}

\begin{table}[h]
\caption{Step-like variations of the specific heat divided by temperature $\Delta C_p^/T$ and of the thermal expansion $\Delta \alpha_a$, $\Delta \alpha_b$, and $\Delta \alpha_c$ at the second-order transition temperatures $T_{HTI,1}$ in Co$_3$V$_2$O$_8$ and $T_{HTI,2}$,  $T_{LTI,2}$, and $T_{C'}$ in Ni$_3$V$_2$O$_8$, and uniaxial pressure dependences of these transitions extracted using the Ehrenfest relation.}
\begin{ruledtabular}
\begin{tabular}{cllllllll}
Ni-content&Transition&$\Delta(C_p/T)$&$\Delta\alpha_a$&$\Delta\alpha_b$&$\Delta\alpha_c$&$\partial(\rm{ln}$~$T_x)/\partial p_a$&$\partial\rm{ln}$~$T_x/\partial p_b$&$\partial\rm{ln}$~$T_x/\partial p_c$\\
$x$&temperature&(mJ/molK)&($10^{-6}$~K$^{-1}$)&($10^{-6}$~K$^{-1}$)&($10^{-6}$~K$^{-1}$)&($10^{-2}$~kbar$^{-1}$)&($10^{-2}$~kbar$^{-1}$)&($10^{-2}$~kbar$^{-1}$)\\
\hline
0 (Co$_3$V$_2$O$_8$)&$T_{HTI,1}$&$480\pm70$&$8.24\pm0.50$&$12.8\pm0.8$&$-12.9\pm2.0$&$1.2\pm0.2$&$1.6\pm0.3$&$-1.6\pm0.4$\\
\hline
&$T_{HTI,2}$&$1005\pm100$&$1.3\pm0.2$&$-9.1\pm0.8$&$11.4\pm1.0$&$0.12\pm0.03$&$-0.8\pm0.02$&$1.0\pm0.2$\\
(Ni$_3$V$_2$O$_8$)&$T_{LTI,2}$&$1080\pm100$&$-3.35\pm0.50$&$-2.25\pm0.25$&$6.49\pm0.50$&$-0.4\pm0.1$&$-0.300\pm0.005$&$0.8\pm0.1$\\
&$T_{C'}$&$881\pm70$&$10\pm1$&$-2.98\pm0.40$&$-9.1\pm1.0$&$3.8\pm0.7$&$-1.1\pm0.2$&$-3.5\pm0.4$\\
\end{tabular}
\end{ruledtabular}
\label{table_second_order}
\end{table}

\end{widetext}

\section{Conclusion} \label{conclusion}

In summary, we have examined in detail the magnetic phase transitions and phase diagram of (Co$_{1-x}$Ni$_x$)$_3$V$_2$O$_8$ using heat capacity and high-resolution thermal expansion measurements. For the parent compounds Co$_3$V$_2$O$_8$ and Ni$_3$V$_2$O$_8$, the same sequences of transitions as that reported in the literature are observed. Our heat capacity data show that there are no further phase transitions below 2~K. There is a smooth evolution of entropy with Ni substitution, going from the full $S=1$ Ni$^{2+}$ magnetic entropy in Ni$_3$V$_2$O$_8$ to half of the $S=3/2$ Co$^{2+}$ magnetic entropy in Co$_3$V$_2$O$_8$. The thermal expansion data together with the heat capacity data allow us to calculate the uniaxial pressure effects on the transition temperatures without actually applying any real pressure. We find that the uniaxial pressure effects for both Co$_3$V$_2$O$_8$ and Ni$_3$V$_2$O$_8$ are extremely anisotropic along the three crystallographic directions. Intriguingly, the anisotropy of Co$_3$V$_2$O$_8$ and Ni$_3$V$_2$O$_8$ are just the opposite, suggesting an intimate connection between the magnetic ordering in both compounds.  This is a little surprising, since the incommensurate antiferromagnetic structure in Co$_3$V$_2$O$_8$ and Ni$_3$V$_2$O$_8$ have propagation vectors along different directions and Co- and Ni- ions have different spin states. Detailed calculations of the magnetic exchange interactions under uniaxial stress are needed to explain this. The thermal expansion data suggest that uniaxial pressure should be a good tuning parameter, i.e., that applying uniaxial pressure along the $c$-axis of Co$_3$V$_2$O$_8$ will lead to a suppression of all phase transitions, whereas $c$-axis pressure in Ni$_3$V$_2$O$_8$ is expected to increase the transition temperatures. It would be particularly interesting to suppress the magnetic ordering in Co$_3$V$_2$O$_8$ to zero temperature by $c$-axis uniaxial pressure and then to see if a different magnetic order emerges at even higher pressures. Unfortunately, one would need about 5 GPa, which is experimentally not very feasible. Closely related to the pressure dependences of the transition temperatures are the magnetic Gr\"{u}neisen parameters, which for Co$_3$V$_2$O$_8$ and Ni$_3$V$_2$O$_8$ exhibit a divergent-like behavior over a limited temperature interval reminiscent of a system close to a quantum critical instability. The anomalous Gr\"{u}neisen parameters are greatly reduced in magnitude for the intermediate concentrations, probably due to disorder effects. However, a small sign change of the Gr\"{u}neisen parameters is observed for intermediate compositions near $x_c\simeq0.25$, which may be the signal of the remnant doping induced transition between the two propagation directions. The cascade of magnetic ordering transitions already pointed to a quite complex ordering scenario in both Co$_3$V$_2$O$_8$ and Ni$_3$V$_2$O$_8$, and the complex behavior of the magnetic Gr\"{u}neisen parameters observed in the present work corroborates this complexity.

\section*{Acknowledgments}

We acknowledge G. Lawes, T. Lorenz and M. Garst for useful discussions and G. Lawes for communicating us data prior to publication.

\appendix

\section{Focus on the transitions}
\label{appendixA}

Here, we detail how we have extracted the characteristic temperatures, as well as the jumps $\Delta S$, $\Delta L_i/L_i$, $\Delta C_p$, and $\Delta \alpha_i$, at the different magnetic phase transitions of all (Co$_{1-x}$Ni$_x$)$_3$V$_2$O$_8$ samples studied in the present work.

Figure \ref{deltaSL} shows how the different first-order transition temperatures have been extracted for each $S$ and $L_i$ data curve, as well as the jumps $\Delta S$ and $\Delta L_i/L_i$ at the transitions. Figure \ref{deltaCpalpha} shows how the different second-order transition temperatures of each sample have been extracted from each of our $C_p$ and $\alpha_i$ data curve, as well as the jumps $\Delta C_p$ and $\Delta \alpha_i$ at the transitions.

Table \ref{table_appendix} presents the values of the different transition temperatures extracted here from each data curve ($C_p$, $\alpha_a$, $\alpha_b$, and $\alpha_c$). In Table \ref{table_transitions}, an average value (with an appropriate error bar) has been retained for each transition.

\begin{widetext}

\begin{figure}[h]
    \centering
    \epsfig{file=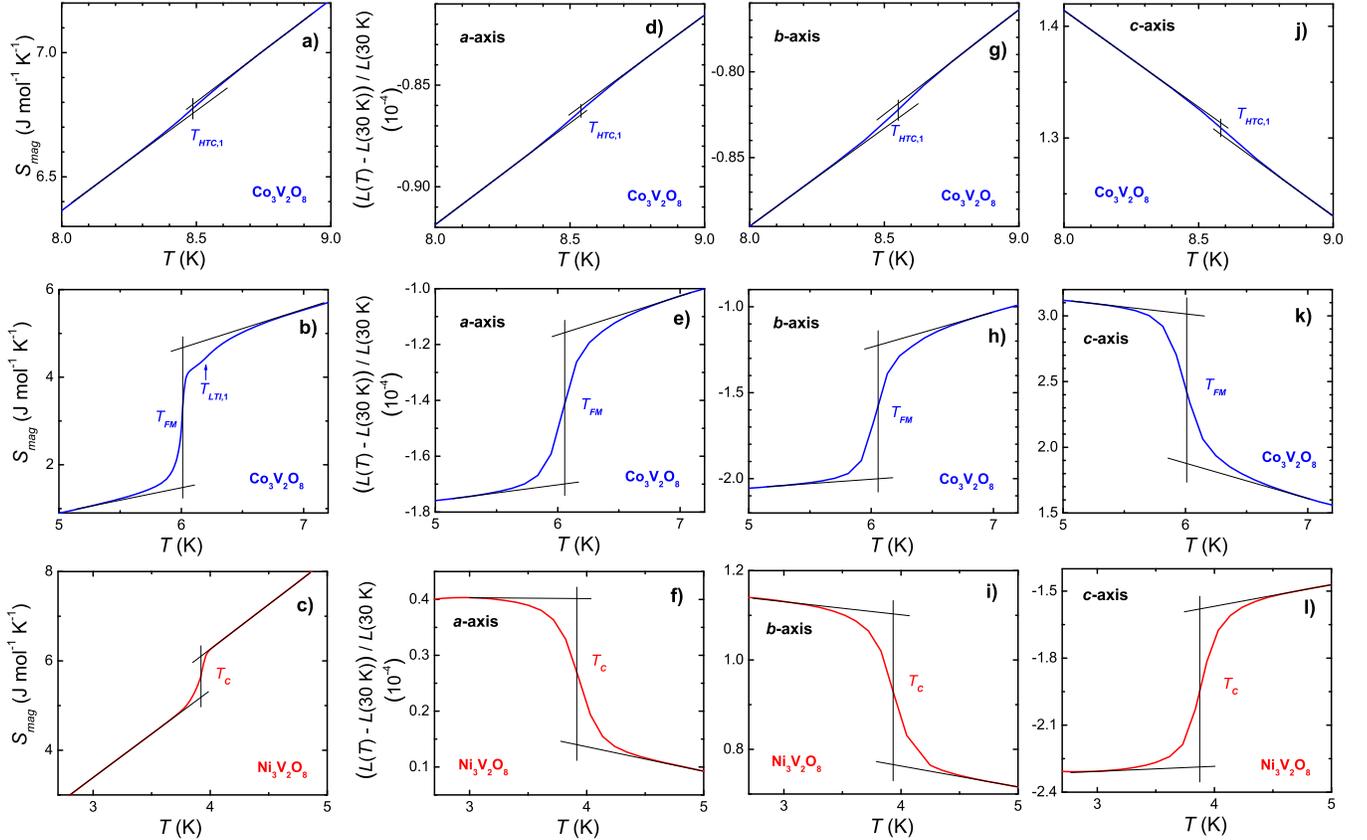,width=180mm}
    \caption{(Color online). Focus on the anomalies at the first-order transitions temperatures in the magnetic entropy and length variations.}
    \label{deltaSL}
\end{figure}

\begin{figure}[h]
    \centering
    \epsfig{file=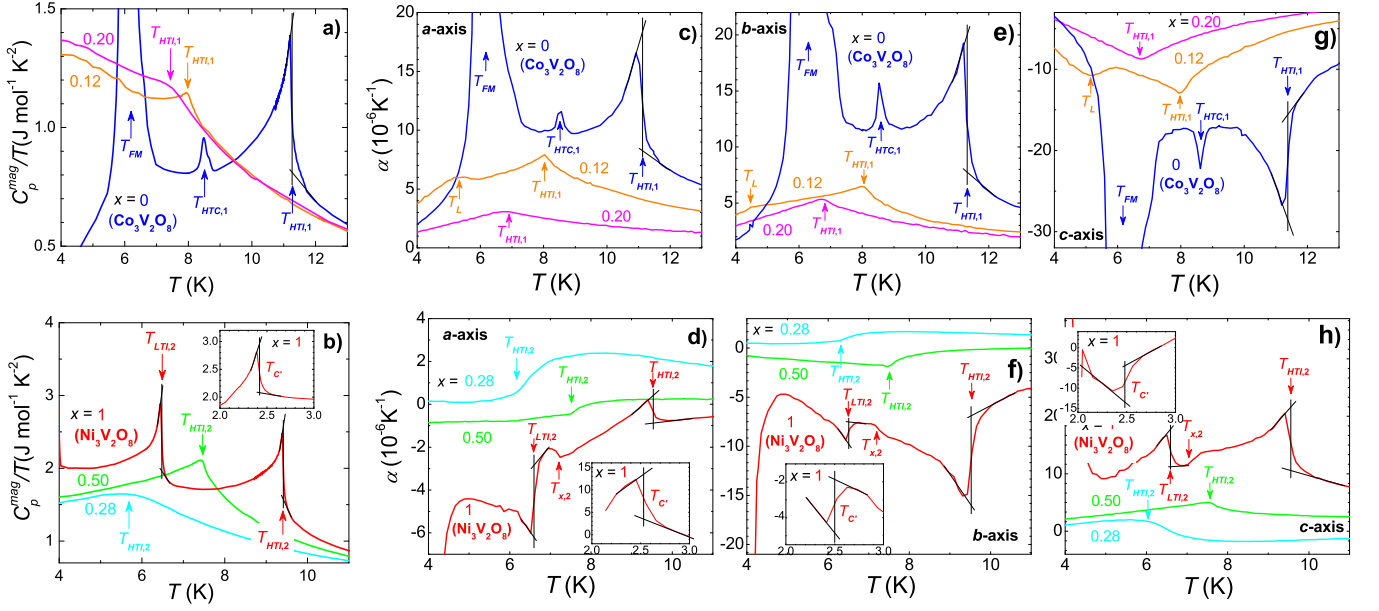,width=180mm}
    \caption{(Color online). Focus on the anomalies at the different second-order transitions temperatures in the heat capacity and thermal expansion.}
    \label{deltaCpalpha}
\end{figure}

\begin{table}[h]
\caption{Transition temperatures extracted here from each data set (specific heat, thermal expansion along $a$, $b$ and $c$, and nature (first- or second-order) of the transitions.}
\begin{ruledtabular}
\begin{tabular}{cllllll}
Ni-content&Transition&Order&from $C_p$&from $\alpha_a$&from $\alpha_b$&from $\alpha_c$\\
$x$&temperature&&\;\;\;(K)&\;\;\;(K)&\;\;\;(K)&\;\;\;(K)\\
\hline
&$T_{HTI,1}$&second&$11.26\pm0.02$&$11.12\pm0.05$&$11.31\pm0.05$&$11.38\pm0.05$\\
0&$T_{HTC,1}$&first&$8.50\pm0.05$&$8.50\pm0.05$&$8.55\pm0.05$&$8.60\pm0.05$\\
(Co$_3$V$_2$O$_8$)&$T_{LTI,1}$&first&$6.21\pm0.05$&-&-&-\\
&$T_{FM}$&first&$6.01\pm0.05$&$6.05\pm0.05$&$6.01\pm0.05$&$6.00\pm0.05$\\
\hline
0.12&$T_{HTI,1}$&second&$8.0\pm0.1$&$8.0\pm0.3$&$8.05\pm0.30$&$8.0\pm0.3$\\
&$T_{L}$&second&-&$5.3\pm0.2$&$4.45\pm0.50$&$5.15\pm0.20$\\
\hline
0.20&$T_{HTI,1}$&second&$7.45\pm0.20$&$6.9\pm0.3$&$6.8\pm0.3$&$6.7\pm0.3$\\
\hline
0.28&$T_{HTI,2}$&second&$5.65\pm0.30$&$6.2\pm0.3$&$6.3\pm0.3$&$6.05\pm0.30$\\
\hline
0.35&$T_{HTI,2}$&second&$6.05\pm0.30$&$6.35\pm0.30$&$6.28\pm0.30$&$6.35\pm0.30$\\
\hline
0.50&$T_{HTI,2}$&second&$7.45\pm0.50$&$7.5\pm0.3$&$7.5\pm0.3$&$7.55\pm0.30$\\
\hline
&$T_{HTI,2}$&second&$9.42\pm0.02$&$9.54\pm0.05$&$9.52\pm0.05$&$9.55\pm0.10$\\
&$T_{x,2}$&second&-&$7.13\pm0.05$&$7.25\pm0.05$&$7.16\pm0.05$\\
1&$T_{LTI,2}$&second&$6.48\pm0.02$&$6.60\pm0.04$&$6.50\pm0.05$&$6.59\pm0.05$\\
(Ni$_3$V$_2$O$_8$)&$T_{C}$&first&$3.93\pm0.02$&$3.93\pm0.05$&$3.94\pm0.05$&$3.90\pm0.05$\\
&$T_{C'}$&second&$2.42\pm0.03$&$2.53\pm0.05$&$2.50\pm0.05$&$2.48\pm0.05$\\
\end{tabular}
\end{ruledtabular}
\label{table_appendix}
\end{table}

\end{widetext}


\begin{thebibliography}{20}
\bibitem{notespinwaves} Far below the magnetic ordering temperature, ungapped ferromagnetic spin waves are expected to lead to a $T^d$ contribution to the specific heat, while ungapped antiferromagnetic spin waves are expected to lead to a $T^{d/2}$ contribution to the specific heat ($d$ is the dimensionality of the magnetic exchange).
\bibitem{note_order_transition} A sharp maximum in $C_p$ is the signature of a first-order phase transition, while a second-order phase transition is generally associated with a step-like or kink-like anomaly in $C_p$.
\end{thebibliography}
\end{document}